# Kernel Bounds for Structural Parameterizations of Pathwidth[*]

Hans L. Bodlaender, Bart M. P. Jansen, and Stefan Kratsch

Utrecht University, The Netherlands,
{h.l.bodlaender,b.m.p.jansen,s.kratsch}@uu.nl

**Abstract.** Assuming the AND-distillation conjecture, the PATHWIDTH problem of determining whether a given graph $G$ has pathwidth at most $k$ admits no polynomial kernelization with respect to $k$. The present work studies the existence of polynomial kernels for PATHWIDTH with respect to other, structural, parameters.

Our main result is that, unless NP $\subseteq$ coNP/poly, PATHWIDTH admits no polynomial kernelization even when parameterized by the vertex deletion distance to a clique, by giving a cross-composition from CUTWIDTH. The cross-composition works also for TREEWIDTH, improving over previous lower bounds by the present authors. For PATHWIDTH, our result rules out polynomial kernels with respect to the distance to various classes of polynomial-time solvable inputs, like interval or cluster graphs.

This leads to the question whether there are nontrivial structural parameters for which PATHWIDTH does admit a polynomial kernelization. To answer this, we give a collection of graph reduction rules that are safe for PATHWIDTH. We analyze the success of these results and obtain polynomial kernelizations with respect to the following parameters: the size of a vertex cover of the graph, the vertex deletion distance to a graph where each connected component is a star, and the vertex deletion distance to a graph where each connected component has at most $c$ vertices.

## 1 Introduction

The notion of kernelization provides a systematic way to mathematically analyze what can be achieved by (polynomial-time) preprocessing of combinatorial problems [12]. This paper discusses kernelization for the problem to determine the *pathwidth* of a graph. The notion of pathwidth was introduced by Robertson and Seymour in their fundamental work on graph minors [16], and is strongly related to the notion of treewidth. There are several notions that are equivalent to pathwidth including *interval thickness*, *vertex separation number*, and *node search number* (see [3] for an overview). The problem to determine the pathwidth of a graph is well studied, also under the different names of the problem.

It is well known that the decision problem corresponding to pathwidth is NP-complete, even on restricted graph classes such as bipartite graphs and chordal graphs [1,13]. A commonly employed practical technique is therefore to preprocess the input before trying to compute the pathwidth, by employing a set of (reversible) data reduction rules. Similar preprocessing techniques for the TREEWIDTH problem have been studied in detail [7,17], and their practical use has been verified in experiments [8]. Using the concept of kernelization we may analyze the quality of such preprocessing procedures within the framework of parameterized complexity. A *parameterized problem* is a language $Q \subseteq \Sigma^* \times \mathbb{N}$, and such a problem is (strongly uniform) *fixed-parameter tractable* (FPT) if there is an algorithm that decides membership of an instance $(x, k)$ in time $f(k)|x|^{\mathcal{O}(1)}$ for some computable function $f$. A *kernelization* (or *kernel*) for $Q$ is a polynomial-time algorithm that transforms each input $(x, k)$ into an *equivalent* instance $(x', k')$ such that $|x'|, k' \leq g(k)$ for some computable function $g$, which is the *size* of the kernel. Kernels of polynomial size are of particular interest due to their practical applications. To analyze the quality of preprocessing rules for PATHWIDTH we therefore study whether they yield polynomial kernels for suitable parameterizations of the PATHWIDTH problem.

---

[*] This work was supported by the Netherlands Organization for Scientific Research (N.W.O.), project "KERNELS: Combinatorial Analysis of Data Reduction".

As the pathwidth of a graph equals the maximum of the pathwidth of its connected components, the PATHWIDTH problem with standard parameterization is AND-compositional and thus has no polynomial kernel unless the AND-distillation conjecture does not hold [4]. We thus do not expect to have kernels for PATHWIDTH of size polynomial in the target value for pathwidth $k$, and we consider whether polynomial kernels can be obtained with respect to other parameterizations.

As PATHWIDTH is known to be polynomial-time solvable when restricted graph classes such as interval graphs [3], trees [11] and cographs [9], it seems reasonable to think that determining the pathwidth of a graph $G$ which is "almost" an interval graph should also be polynomial-time solvable. Formalizing the notion of "almost" as the number of vertices that have to be deleted to obtain a graph in the restricted class $\mathcal{F}$, we can study the extent to which data reduction is possible for graphs which are close to polynomial-time solvable instances through the following problem:

> PATHWIDTH PARAMETERIZED BY A MODULATOR TO $\mathcal{F}$
> **Instance:** A graph $G = (V, E)$, a positive integer $k$, and a set $S \subseteq V$ such that $G - S \in \mathcal{F}$.
> **Parameter:** $\ell := |S|$.
> **Question:** $\mathbf{pw}(G) \leq k$?

The set $S$ is a *modulator* to the class $\mathcal{F}$. Observe that pathwidth should be polynomial-time solvable on $\mathcal{F}$ in order for this parameterized problem to be FPT. Our main result is a kernel lower bound for such a parameterization of PATHWIDTH. We prove that despite the fact that the pathwidth of an interval graph is simply the size of its largest clique minus one — which is very easy to find on interval graphs — the PATHWIDTH problem parameterized by a modulator to an interval graph does not admit a polynomial kernel unless NP $\subseteq$ coNP/poly. In fact, we prove the stronger statement that, under the same condition, PATHWIDTH parameterized by a modulator to a single clique (i.e., by distance to $\mathcal{F}$ consisting of all complete graphs) does not admit a polynomial kernel[1] (Section 5). As the graph resulting from the lower-bound construction is co-bipartite, its pathwidth and treewidth coincide [14]: a corollary to our theorem therefore shows that TREEWIDTH parameterized by vertex-deletion distance to a clique does not admit a polynomial kernel unless NP $\subseteq$ coNP/poly, thereby significantly strengthening a result of our earlier work [7] where we only managed to prove kernel lower bounds by modulators from cluster graphs and co-cluster graphs.

Our kernel bound effectively shows that even in graphs which are cliques after the deletion of $k$ vertices, the information contained in the (non)edges between these $k$ vertices and the clique is such that we cannot decrease the size of the clique to polynomial in $k$ in polynomial time, without changing the answer in some cases.

Faced with these negative results, we try to formulate *safe* reduction rules for PATHWIDTH (Section 3). It turns out that many of the rules for TREEWIDTH (e.g., the rules involving (almost) simplicial vertices) are invalid when applied to PATHWIDTH, and more careful reduction procedures are needed to reduce the number of such vertices. We obtain several reduction rules for pathwidth, and show that they lead to provable data reduction guarantees when analyzed using a suitable parameterization (Section 4). In particular we prove that PATHWIDTH parameterized by a vertex cover $S$ (i.e., using $\mathcal{F}$ as the class of edgeless graphs in the template above) admits a kernel with $\mathcal{O}(|S|^3)$ vertices, that the parameterization by a modulator $S'$ to a disjoint union of stars has a kernel with $\mathcal{O}(|S'|^4)$ vertices, and finally that parameterizing by a set $S''$ whose deletion leaves a graph in which every connected component has at most $c$ vertices admits a kernel with $\mathcal{O}(c \cdot |S''|^3 + c^2 \cdot |S''|^2)$ vertices.

## 2 Preliminaries

In this work all graphs are finite, simple, and undirected. The open neighborhood of a vertex $v \in V$ in a graph $G$ is denoted by $N_G(v)$, and its closed neighborhood is $N_G[v]$. For sets of vertices $W \subseteq V$ we let $N_G[W] = \bigcup_{v \in W} N_G[v]$ and $N_G(W) = N_G[W] \setminus W$. If $S \subseteq V$ is a vertex set then $G - S$ denotes the graph obtained from $G$ by deleting all vertices of $S$ and their incident edges. For a single vertex $v$ we write $G - v$ instead of $G - \{v\}$. A vertex $v$ is *simplicial* in a graph $G$ if $N_G(v)$ is a clique. A vertex $v \in V$ is *almost simplicial* in a graph $G$ if $v$ has a neighbor $w$ such that $N_G(v) \setminus \{w\}$ is a clique. In such a case, we call $w$ the *special neighbor* of $v$. For a set of vertices $W \subseteq V$, the subgraph of $G$ induced by $W$ is denoted as $G[W]$. A *path decomposition* of a graph $G = (V, E)$ is a non-empty sequence $(X_1, \ldots, X_r)$ of subsets of $V$ called *bags*, such that:

---

[1] For completeness we point out that PATHWIDTH parameterized by a modulator to a clique is FPT: try all orderings in which the vertices from $S$ can be introduced and forgotten in a decomposition, and do a polynomial-time computation for each ordering to find the best way to fit the clique $G - X$ into the decomposition.

- $\bigcup_{1 \leq i \leq r} X_i = V$,
- for all edges $\{v, w\} \in E$ there is a bag $X_i$ containing $v$ and $w$, and
- for all vertices $v \in V$, the bags containing $v$ are consecutive in the sequence.

The *width* of a path decomposition is $\max_{1 \leq i \leq r} |X_i| - 1$. The *pathwidth* $\mathbf{pw}(G)$ of $G$ is the minimum width of a path decomposition of $G$. Throughout the paper we will often make use of the fact that the pathwidth of a graph does not increase when taking a minor. We also use the following results.

**Lemma 1 (Cf. [9]).** *If graph $G$ contains a clique $W$ then any path- or tree decomposition for $G$ has a bag containing all vertices of $W$.*

**Lemma 2.** *All graphs $G$ admit a minimum-width path decomposition in which each simplicial vertex is contained in exactly one bag of the decomposition.*

*Proof.* Lemma 1 shows that for each simplicial vertex $v$, any path decomposition of $G$ has a bag containing the clique $N[v]$. As removal of $v$ from all other bags preserves the validity of the decomposition, we may do so successively for all simplicial vertices to obtain a decomposition of the desired form. □

## 3   Reduction Rules

In this section we give a collection of reduction rules. Formally, each rule takes as input an instance $(G, S, k)$ of PATHWIDTH PARAMETERIZED BY A MODULATOR TO $\mathcal{F}$, and outputs an instance $(G', S', k')$. With the exception of occasionally outright deciding **yes** or **no**, none of our reduction rules change the modulator $S$ or the value of $k$. In the interest of readability we shall therefore be less formal in our exposition, and make no mention of the values of $S'$ and $k'$ in the remainder; they will be understood to be equal to $S$ and $k$.

We say that a rule is *safe for pathwidth* (or in short: safe) if for each input $(G, S, k)$ and output $(G', S', k')$, the pathwidth of $G$ is at most $k$ if and only if the pathwidth of $G'$ is at most $k'$. Any subset of the rules gives a 'safe' preprocessing algorithm for pathwidth: apply the rules until no longer possible. We will argue later that this takes polynomial time for our rules, and give kernel bounds for some parameters of the graphs.

### 3.1   Vertices of small degree

We start off with a few simple rules for vertices of small degree. Note that, necessarily, these rules are slightly more restrictive than for the treewidth case; e.g., we cannot simply delete vertices of degree one since trees have treewidth one but unbounded pathwidth. The first rule is trivial.

**Rule 1** *Delete any vertex of degree zero.*

**Rule 2** *If two degree-one vertices share their neighbor then delete one of them.*

Correctness of Rule 2 follows from insights on the pathwidth of trees, pioneered by Ellis et al. [11]. A self-contained proof is provided in the appendix.

The following rule handles certain vertices of degree two; a correctness proof is given in the appendix.

**Rule 3** *Let $v, w$ be two vertices of degree two, and suppose $x$ and $y$ are common neighbors to $v$ and $w$ with $x \in S$. Then remove $w$ and add the edge $\{x, y\}$.*

The differences with safe rules for treewidth are interesting to note: for treewidth, we can remove vertices of degree one, and remove vertices of degree two when adding an edge between their neighbors.

### 3.2 Common neighbors and disjoint paths

Rule 4 in this section also appears in our work on kernelization for treewidth [7] and traces back to well-known facts about treewidth (e.g.,[2,10]). It is also safe in the context of pathwidth; the safeness proof is identical to when dealing with treewidth and is hence deferred to the appendix.

**Lemma 3.** *Let $v$ and $w$ be nonadjacent vertices. Suppose there are at least $k+1$ internally vertex disjoint paths from $v$ to $w$ in $(V, E)$. Then the pathwidth of $G$ is at most $k$, if and only if the pathwidth of $G' = (V, E \cup \{\{v, w\}\})$ is at most $k$.*

A special case of Lemma 3, and the implied Rule 4, is when $v$ and $w$ have at least $k + 1$ common neighbors. As we do not want to increase the size of a modulator, we only add edges between pairs of vertices with at least one endpoint in the modulator; thus $G - S$ remains unchanged.

**Rule 4 (Disjoint paths (with a modulator))** *Let $v \in S$ be nonadjacent to $w \in V$, and suppose there are at least $k+1$ paths from $v$ to $w$ that only intersect at $v$ and $w$, where $k$ denotes the target pathwidth. Then add the edge $\{v, w\}$.*

### 3.3 Simplicial vertices

In this section, we give a safe rule that helps to bound the number of simplicial vertices of degree at least two in a graph. Recall that we already have rules for vertices of degree one and zero, which are trivially simplicial.

**Lemma 4.** *Let $G = (V, E)$ be a graph, and let $v \in V$ be a simplicial vertex of degree at least two. If for all $x, y \in N_G(v)$ with $x \neq y$ there is a simplicial vertex $w \notin N_G[v]$ such that $x, y \in N_G(w)$, then $\mathbf{pw}(G) = \mathbf{pw}(G - v)$.*

*Proof.* As $G - v$ is a subgraph of $G$, we directly have that $\mathbf{pw}(G - v) \leq \mathbf{pw}(G)$. For the converse, let $(X_1, \ldots, X_r)$ be an optimal path decomposition of $G - v$. Using Lemma 2, we assume that for each simplicial vertex $x$, there is a unique bag $X_{i_x}$ with $N_G[x] \subseteq X_{i_x}$.

Let $C = N_G(v)$. A bag that contains $C$ is called a $C$-bag. As $C$ is a clique, Lemma 1 shows there is at least one $C$-bag. The $C$-bags must be consecutive in the path decomposition; let them be $X_{i_1}, \ldots, X_{i_2}$. We will first show there is a vertex $w \notin N_G[v]$ which is simplicial in $G - v$, and is contained in a $C$-bag. Let $x, y \in C$ (possibly with $x = y$) be vertices such that $x$ does not occur in bags with index smaller than $i_1$, and $y$ does not occur in bags of index larger than $i_2$.

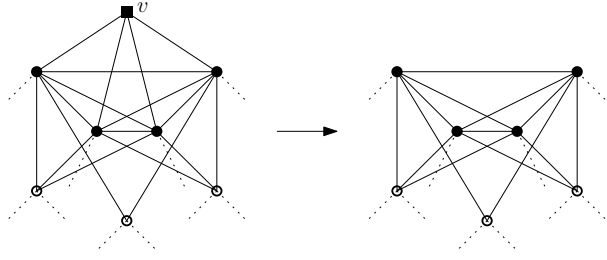

**Fig. 1.** An example of an application of the Simplicial vertex rule.

If $x \neq y$ then let $w \notin N_G[v]$ be simplicial in $G$ such that $x, y \in N_G(w)$, whose existence is guaranteed by the preconditions. As $w$ is also simplicial in $G - v$ it occurs in a unique bag, which must be a $C$-bag since it must meet its neighbors $x$ and $y$ there. If $x = y$ then, as $v$ has degree at least two, there is a vertex $w \notin N_G[v]$ which is simplicial in $G$ and adjacent to $x$; hence its unique occurrence is also in a $C$-bag.

Thus we have established there is a vertex $w \notin N_G[v]$ which is simplicial in $G - v$ and is contained in exactly one bag, which is a $C$-bag $X_i$. Now insert a new bag just after $X_i$, with vertex set $(X_i \setminus \{w\}) \cup \{v\}$. As $X_i \setminus \{w\}$ contains all $v$'s neighbors, this gives a path decomposition of $G$ without increasing the width, and concludes the proof. □

Lemma 4 directly shows that Rule 5 is safe for PATHWIDTH.

**Rule 5** *Let $v$ be a simplicial vertex of degree at least two. If for all $x, y \in N_G(v)$ with $x \neq y$ there is a simplicial vertex $w \notin N_G[v]$ such that $x, y \in N_G(w)$, then remove $v$.*

### 3.4   Simplicial components

Let $S$ be the set of vertices used as the modulator. We say that a set of vertices $W$ is a *simplicial component* if $W$ is a connected component in $G - S$ and $N_G(W) \cap S$ is a clique. Our next rule deals with simplicial components.

**Rule 6 (Simplicial components of known pathwidth)** *Let $S \subseteq V$ be the modulator and let $k$ denote the target pathwidth. Suppose that for each pair $v, w \in S \cap N_G(W)$ (including $v = w$), there are at least $2k+3$ simplicial components $Z \neq W$ such that $\{v, w\} \subseteq N_G(Z)$ and $\mathbf{pw}(G[Z]) \geq \mathbf{pw}(G[W])$. Then remove $W$ and its incident edges.*

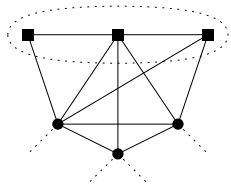

**Fig. 2.** The vertices marked with a square box form a simplicial component when interpreting the round vertices as the modulator.

Note that we have to include the case $v = w$ to ensure correctness for simplicial components which are adjacent to exactly one vertex in the modulator. Lemma 9 in the appendix shows that Rule 6 is safe.

Let us briefly discuss the running time of this reduction rule. As the modulator ensures that $G - S$ is contained in the graph class $\mathcal{F}$, the rule can be applied in polynomial time if the pathwidth of graphs in $\mathcal{F}$ can be determined efficiently. In the setting in which we apply the rule, the graphs in $\mathcal{F}$ are either disjoint unions of stars (which are restricted types of forests, allowing the use of the linear-time algorithm of Ellis et al. [11]), or $\mathcal{F}$ has constant pathwidth which means that the FPT algorithm for $k$-PATHWIDTH [2] runs in linear time.

### 3.5 Almost simplicial vertices

For almost simplicial vertices, we have a rule that replaces an almost simplicial vertex by a number of vertices of degree two. In several practical settings, the increase of number of vertices may be undesirable; the rule is useful to derive some theoretical bounds.

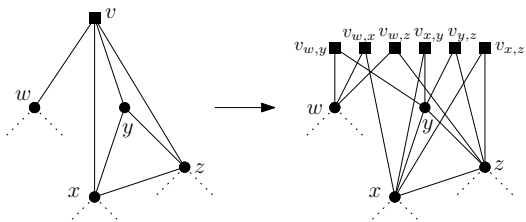

**Fig. 3.** An example of an application of the rule for almost simplicial vertices.

**Lemma 5.** *Let $G = (V, E)$ be a graph and let $v \in V$ be an almost simplicial vertex of degree at least three, with special neighbor $w$. Let $G'$ be obtained by deleting $v$ and by adding a vertex $v_{p,q}$ with neighbors $p$ and $q$ for any $p, q \in N_G(v)$ with $p \neq q$. Then $\mathbf{pw}(G) = \mathbf{pw}(G')$.*

The proof of the lemma is moved to the appendix. The lemma justifies the following reduction rule, by observing that an almost simplicial vertex $v$ with $\deg_G(v) > k+1$ means that $\mathbf{pw}(G) > k$, as $N_G[v] - w$ then forms a clique of size at least $k+2$.

**Rule 7** *Let $v \in V \setminus S$ be an almost simplicial vertex of degree at least three with special neighbor $w$. Let $k$ be the target pathwidth. If $\deg_G(v) > k+1$ then output* **no***. Otherwise, delete $v$ and add a vertex $v_{p,q}$ with neighbors $p$ and $q$ for any $p, q \in N(v)$ with $p \neq q$.*

As a simplicial vertex is trivially almost simplicial, note that — in comparison to Rule 5 — the previous rule gives an alternative way of dealing with simplicial vertices.

## 4 Polynomial kernelizations

For each of the safe rules given in the previous section, there is a polynomial time algorithm that tests if the rule can be applied, and if so, modifies the graph accordingly. (We assume that for Rule 6 the bound $\ell$ on the pathwidth of the components is a constant.) The following lemma shows that any algorithm that exhaustively applies (possibly just a subset of) these reduction rules can be implemented to run in polynomial time.

**Lemma 6.** *Each input instance $(G, S, k)$ is exhaustively reduced by $\mathcal{O}(n^2 + nk^2)$ applications of the reduction rules.*

*Proof.* First we note that for non-trivial instances, Rule 4 does not add edges to a vertex of degree at most two. In particular, no rule increases the number of vertices of degree at least three. So, we have at most $n$ applications of a rule that removes a vertex of degree at least three, and $\mathcal{O}(n^2)$ applications of Rule 4. Rule 7 is therefore executed at most $n$ times in total, and thus the number of vertices of degree two that are added in these steps is bounded by $\mathcal{O}(nk^2)$. As each other rule removes at least one vertex, the total number of rule applications in $G$ is bounded by $\mathcal{O}(n^2 + nk^2)$. □

By analyzing our reduction rules with respect to different structural parameters, we get the following results.

**Theorem 1.** PATHWIDTH PARAMETERIZED BY A MODULATOR TO $\mathcal{F}$ *admits polynomial kernels for the following choices of $\mathcal{F}$:*

1. *A kernel with $\mathcal{O}(\ell^3)$ vertices when $\mathcal{F}$ is the class of all independent sets, i.e., if the modulator $S$ is a vertex cover.*
2. *A kernel with $\mathcal{O}(c \cdot \ell^3 + c^2 \cdot \ell^2)$ vertices when $\mathcal{F}$ is the class of all graphs with connected components of size at most $c$.*
3. *A kernel with $\mathcal{O}(\ell^4)$ vertices when $\mathcal{F}$ is the class of all disjoint unions of stars.*

*Proof.* We show Part 3 followed by Part 2. Part 1 follows from the latter since it is a special case corresponding to $c = 1$.

**(Part 3.)** As stars have pathwidth one, graphs with a modulator $S$ of size $\ell$ to a set of stars have pathwidth at most $\ell + 1$. Thus, if $k \geq \ell + 1$, we return a dummy **yes**-instance of constant size. Now, assume $k \leq \ell$.

Our kernelization applies Rules 1–6 while possible, and applies Rule 7 to all vertices which have at most one neighbor in $G - S$. (Applying the rule to vertices with more neighbors in $G - S$ might cause the resulting graph $G' - S$ not to be a disjoint union of stars.) Recall for Rule 6 that $\mathbf{pw}(G - S) \leq 1$.

Let $(G, S, k)$ be a reduced instance. We will first bound the number of connected components of $G - S$, with separate arguments for simplicial and nonsimplicial components. Each component is a star, i.e., it is a single vertex or a $K_{1,r}$ for some $r$ (a center vertex with $r$ leaves). Note that in this proof the term leaf refers to a leaf of a star in $G - S$, independent of its degree in $G$ (and all degrees mentioned are with respect to $G$).

Associate each nonsimplicial component $C$ of $G - S$ to an arbitrary pair of nonadjacent neighbors of $C$ in $S$. It is easy to see that each such component provides a path between the two chosen neighbors, and that for different components these paths are internally vertex disjoint. Thus, since Rule 4 does not apply, no pair of vertices of $S$ has more than $k$ components associated to it. Hence there are at most $k \cdot |S|^2 = \mathcal{O}(\ell^3)$ nonsimplicial components.

Now consider a simplicial component $W$ of $G - S$, and note that $\mathbf{pw}(G[W]) \leq 1$. As Rule 6 does not apply, there is a pair $v, w \in S \cap N_G(W)$ (possibly $v = w$) such that there are strictly less than $2k + 3$ simplicial components $W' \neq W$ with $\mathbf{pw}(G[W']) \geq \mathbf{pw}(G[W])$ and $\{v, w\} \subseteq N_G(W')$. Associate $W$ to the pair $v, w$. It follows immediately that no pair of vertices of $S$ has more than $2k + 3$ components associated to it, which gives a bound of $(2k + 3) \cdot |S|^2 = \mathcal{O}(\ell^3)$ on the number of simplicial components.

Thus we find that $G - S$ has a total of $\mathcal{O}(\ell^3)$ connected components (each of which is a star). This bounds the number of centers of stars by $\mathcal{O}(\ell^3)$. It remains to bound the total number of leaves that are adjacent to those centers.

Clearly, each star center has at most one leaf which has degree one (in $G$) by Rule 2. Each leaf of degree two has exactly one neighbor in $S$ in addition to its adjacent star center. Since Rule 3 does not apply, no two leaves of degree two can have the same star center and neighbor in $S$; thus there are at most $\mathcal{O}(\ell^4)$ leaves of degree two.

Now, we are going to count the number of leaves (of stars) that are of degree more than two. For each such leaf, one neighbor is the center of its star and all other neighbors are in $S$. If its neighbors in $S$ would form a clique, then the leaf would be almost simplicial in $G$ (with the star center as the special neighbor) and Rule 7 would apply. Hence, as $G$ is reduced, we can associate each such leaf to a nonadjacent pair of vertices in $S$. As Rule 4 cannot be applied, we associate $\mathcal{O}(k)$ vertices to a pair, and thus the number of such leaves is bounded by $\mathcal{O}(k \cdot \ell^2) = \mathcal{O}(\ell^3)$.

Thus, the total number of vertices in $G$ is bounded by $\mathcal{O}(\ell^4)$. By Lemma 6 the reduction rules can exhaustively be applied in polynomial time. As the rules preserve the fact that $G - S$ is a disjoint union of stars, the resulting instance is a correct output for a kernelization algorithm. This completes the proof of Part 3.

**(Part 2.)** Fix some constant $c$ and let $\mathcal{F}$ be the class of all graphs of component size at most $c$. Let $(G, S, k)$ be an input instance. Note that the pathwidth of $G$ is bounded by $c + |S| - 1$, since each component of $G - S$ has pathwidth at most $c - 1$. We assume that $k \leq c + |S| - 2$; otherwise the instance is **yes** and we may return a dummy **yes**-instance of constant size.

Our algorithm uses Rules 1, 2, 4, 5, and 6. Regarding Rule 6 we note that the pathwidth of components of $G - S$ can be computed in constant time (depending only on $c$).

Consider a graph $G$ where none of these rules can be applied. The bounds for the number of simplicial and nonsimplicial components of $G-S$ work analogously to Part 3; there are $\mathcal{O}(k|S|^2)$ components of the respective types. This gives a total of $\mathcal{O}(|S| + c \cdot (c + |S|) \cdot |S|^2) = \mathcal{O}(c^2|S|^2 + c|S|^3)$ vertices in $G$, using that $k \leq c + |S| - 2$. This completes the proof of Part 2. $\square$

We remark that while Rule 5 is not needed to establish the kernel bounds presented in the previous theorem, we have included it in our presentation for two reasons. First of all, applying the rule may prove to be useful in practical situations. It also leads to improved theoretical bounds on some quantities. For example, when $\mathcal{F}$ is the class of all independent sets (and the modulator $S$ is a vertex cover), the use of Rule 5 will decrease the number of simplicial vertices in $G - S$ to $\mathcal{O}(|S|^2)$, whereas the bound would be $\mathcal{O}(|S|^3)$ without this rule. Alas, as the kernel size for this choice of $\mathcal{F}$ is dominated by the $\Theta(|S|^3)$ term for the number of non-simplicial vertices, we still end up with a cubic-vertex kernel.

## 5 Lower bounds: Modulator to a Single Clique

We complement the positive results of the previous section by some negative results. In particular, we show that the problems TREEWIDTH PARAMETERIZED BY A MODULATOR TO A SINGLE CLIQUE (TWMSC) and PATHWIDTH PARAMETERIZED BY A MODULATOR TO A SINGLE CLIQUE (PWMSC) do not admit a polynomial kernel unless NP $\subseteq$ coNP/poly. In fact, we show that the results hold when restricted to co-bipartite graphs; as for these graphs the pathwidth equals the treewidth [14], the same proof works for both problems. The problems are covered by the general template given in the introduction, when using $\mathcal{F}$ as the class of all cliques. Observe that $\mathcal{F}$ only contains *connected* graphs, and in particular $\mathcal{F}$ is not closed under disjoint union.

To prove the lower bound we employ the technique of cross-composition [6], starting from the following NP-complete version [15, Corollary 2.10] of the CUTWIDTH problem:

CUTWIDTH ON CUBIC GRAPHS (CUTWIDTH3)
**Instance:** A graph $G$ on $n$ vertices in which each vertex has degree at least one and at most three, and an integer $k \leq |E(G)|$.
**Question:** Is there a linear layout of $G$ of cutwidth at most $k$, i.e., a permutation $\pi$ of $V(G)$ such that $\max_{i=1}^{n} |\{\{u,v\} \in E(G) \mid \pi(u) \leq i < \pi(v)\}| \leq k$?

As space restrictions prohibit us from presenting the full proof in this extended abstract, we will sketch the main ideas. To obtain a kernel lower bound through cross-composition, we have to embed the logical OR of a series of $t$ input instances of CUTWIDTH3 on $n$ vertices each into a single instance of the target problem for a parameter value polynomial in $n + \log t$. At the heart of our construction lies an idea of Arnborg et al. [1] employed in their NP-completeness proof for TREEWIDTH. They interpreted the treewidth of a graph as the minimum cost of an elimination ordering on its vertices[2], and showed how for a given graph $G$ a co-bipartite graph $G^*$ can be created such that the cost of elimination orderings on $G^*$ corresponds to the cutwidth of $G$ under a related ordering.

We extend their construction significantly. By the degree bound, instances with $n$ vertices have $\mathcal{O}(n^2)$ different degree sequences. The framework of cross-composition thus allows us to work on instances with the same degree sequence (and same $k$). By enforcing that the structure of one side of the co-bipartite graph $G^*$ only has to depend on this sequence, all inputs can share the same "right hand side" of the co-bipartite graph; this part will remain small and act as the modulator. By a careful balancing act of weight values we then ensure that the cost of elimination orderings on the constructed graph $G^*$ are dominated by eliminating the vertices corresponding to exactly one of the input instances, ensuring that a sufficiently low treewidth is already achieved when one of the input instances is **yes**. On the other hand, the use of a binary-encoding representation of instance numbers ensures that low-cost elimination orderings for $G^*$ do not mix vertices corresponding to different input instances. The remaining details can be found in Appendix D. Our construction yields the following results.

**Theorem 2.** *Unless $NP \subseteq coNP/poly$,* PATHWIDTH *and* TREEWIDTH *do not admit polynomial kernels when parameterized by a modulator to a single clique.*

Interestingly, the parameter at hand is nothing else than the size of a vertex cover in the complement graph.

## 6  Conclusions

In this paper, we investigated the existence of polynomial kernelizations for PATHWIDTH. Taking into account that the problem is already known to be AND-compositional with respect to the target pathwidth — thus excluding polynomial

---

[2] To eliminate a vertex in a graph means to remove it while completing its open neighborhood into a clique. When eliminating the vertices of a graph in the order given by $\pi$, the cost of the elimination ordering $\pi$ is the maximum degree of a vertex at the time it is eliminated.

kernels under the AND-distillation conjecture — we study alternative, structural parameterizations.

Our main result is that PATHWIDTH admits no polynomial kernelization with respect to the number of vertex deletions necessary to obtain a clique, unless NP $\subseteq$ coNP/poly. This rules out polynomial kernels for vertex deletion distance from various interesting graph classes on which PATHWIDTH is known to be polynomial-time solvable, like chordal and interval graphs.

On the positive side we develop a collection of safe reduction rules for PATHWIDTH. Analyzing the effect of the rules we show that they give polynomial kernels with respect to the following parameters: vertex cover (i.e., distance from the class of independent sets), distance from graphs of bounded component size, and distance from disjoint union of stars.

It is an interesting open problem to determine whether there is a polynomial kernel for PATHWIDTH parameterized by the size of a feedback vertex set. For the related TREEWIDTH problem, a kernel with $\mathcal{O}(|S|^4)$ vertices is known [7], where $S$ denotes a feedback vertex set. Regarding PATHWIDTH, long paths in $G - S$ are the main obstacle that needs to be addressed by additional reduction rules.

## A  Safeness of low degree rules

**Lemma 7.** *Rule 2 is safe.*

*Proof.* Let $u$ and $v$ be two vertices of degree one with shared neighbor $w$ and let $G'$ be obtained from $G$ by removing $v$.

By Lemma 2 there is a minimum-width path decomposition $(X_1, \ldots, X_r)$ of $G'$ in which $u$ occurs in a unique bag $X_i$, which must therefore also contain $w$. We obtain a path decomposition for $G'$ by adding a bag $X_{i'}$ containing $(X_i \setminus \{u\}) \cup \{v\}$ next to bag $X_i$. As this does not increase the width, safeness of the rule follows. □

**Lemma 8.** *Rule 3 is safe.*

*Proof.* Suppose we obtain $G'$ from $G$ by applying Rule 3. The graph $G'$ can be obtained from $G$ by contracting the edge $\{w, y\}$ and thus $G'$ is a minor of $G$. Hence the pathwidth of $G'$ is at most the pathwidth of $G$.

As vertex $v$ is simplicial in $G'$, by Lemma 2 we can assume that we have a path decomposition of optimal width containing a unique bag $X_i$ with $v$, $x$ and $y$. Take a new bag $X_{i'}$ with $X_{i'} = X_i - \{v\} \cup \{w\}$, and insert it in the path decomposition directly after $X_i$. This gives a path decomposition of $G$ of the same width. □

## B  Safeness of the simplicial components rule

**Lemma 9.** *Let $k$ be an integer. Let $G = (V, E)$ be a graph with modulator $S \subseteq V$. Suppose that $W \subseteq V - S$ is a simplicial component such that, for each pair of vertices $v, w \in N_G(W) \cap S$ (including $v = w$), there are at least $2k + 3$ simplicial components $Z \neq W$ such that $\mathbf{pw}(G[Z]) \geq \mathbf{pw}(G[W])$ and $\{v, w\} \subseteq N_G(Z)$. Then $\mathbf{pw}(G) \leq k$ if and only if $\mathbf{pw}(G - W) \leq k$.*

*Proof.* Clearly, $\mathbf{pw}(G) \leq k$ implies $\mathbf{pw}(G - W) \leq k$.

Now, suppose that the pathwidth of $G - W$ is at most $k$. Consider a path decomposition $(X_1, \ldots, X_r)$ of $G - W$ of width at most $k$. We assume that $X_1 = X_r = \emptyset$ for notational convenience.

Let $Y = N(W) \cap S$. If $Y = \emptyset$ then $W$ forms a connected component of $G$, and the precondition to the lemma ensures that $\mathbf{pw}(G[W]) \leq \mathbf{pw}(G - W)$ because at least one simplicial component $Z$ with $\mathbf{pw}(G[Z]) \geq \mathbf{pw}(G[W])$ remains in $G - W$; hence adding an optimal path decomposition of $G[W]$ to the decomposition of $G - W$ does not increase its width, as there is no interaction between the two parts. In the remainder we may therefore assume that $Y \neq \emptyset$.

Note that $Y$ is a clique in $G$, so by Lemma 1 and the convexity property of path decompositions there are $j_1 \leq j_2$, such that $Y \subseteq X_j \Leftrightarrow j_1 \leq j \leq j_2$, i.e., the bags that contain all vertices in $Y$ are precisely those with index in $j_1, \ldots, j_2$. We say that a simplicial component $Z$ is *internal* to $Y$ if vertices $v \in Z$ occur only in bags $X_i$ with $j_1 \leq i \leq j_2$.

*Claim.* There is a simplicial component $Z$ in $G - W$ such that $\mathbf{pw}(G[Z]) \geq \mathbf{pw}(G[W])$ and $Z$ is internal to $Y$.

*Proof.* Let $v, w \in N(W) \cap S$ with $v \notin X_{j_1-1}$ and $w \notin X_{j_2+1}$ (possibly $v = w$). By assumption, there are at least $2k + 3$ simplicial components $C$ such that $\mathbf{pw}(G[C]) \geq \mathbf{pw}(G[W])$ and $\{v, w\} \subseteq N(C)$. If one of these is internal to $Y$, then we are done. Consider such a component $C$. As it is not internal, there must be a bag $X_{i_1}$ with $X_{i_1} \cap C \neq \emptyset$ and $i_1 < j_1$ or $j_2 < i_1$.

Suppose $i_1 < j_1$. By adjacency of $C$ to $v$, there must be a bag $i_2$ with $v \in X_{i_2}$ and $X_{i_2} \cap C \neq \emptyset$. As $i_2 \geq j_1$ by choice of $v$, we have established that $C \cap X_{i_1} \neq \emptyset$, $C \cap X_{i_2} \neq \emptyset$ and $i_1 < j_1 \leq i_2$. By connectivity of $C$ there is a path from a vertex in $C \cap X_{i_1}$ to a vertex in $C \cap X_{i_2}$ in $G[C]$. By tracing this path in the path decomposition, it is easy to see that $C \cap X_{j_1} \neq \emptyset$. A similar argument shows that when $j_2 < i_1$ then $C \cap X_{j_2} \neq \emptyset$.

Therefore, each of the $2k + 3$ simplicial components $C$ has a vertex in $X_{j_1}$ or in $X_{j_2}$. So $|X_{j_1}| \geq k + 2$ or $|X_{j_2}| \geq k + 2$, which contradicts the assumption that we had a path decomposition of width $k$. □

Now, let $Z$ be an internal simplicial component of $G-W$ such that $\mathbf{pw}(G[Z]) \geq \mathbf{pw}(G[W])$. Suppose the pathwidth of $G[Z]$ is $\ell$. There must be a bag $X_{i'}$ that contains at least $\ell + 1$ vertices of $Z$, otherwise $(X_1 \cap Z, \ldots, X_r \cap Z)$ is a path decomposition of $Z$ of width less than $\ell$. Let $(Z_1, \ldots, Z_q)$ be a path decomposition of $G[Z]$ of width $\ell$, and let $(W_1, \ldots, W_p)$ be a path decomposition of $G[W]$ of width at most $\ell$.

Now, the following tuple constitutes a path decomposition for $G$ of width at most $k$:

$$(X_1 \setminus Z, \ldots, X_{i'-1} \setminus Z, (X_{i'} \setminus Z) \cup Z_1, \ldots, (X_{i'} \setminus Z) \cup Z_q,$$
$$(X_{i'} \setminus Z) \cup W_1, \ldots, (X_{i'} \setminus Z) \cup W_p, X_{i'+1} \setminus Z, \ldots, X_r).$$

It is obtained by first removing $Z$ from all bags, making $p + q$ copies of $X_{i'}$, and then adding in successive copies a path decomposition of $G[Z]$ followed by a path decomposition of $G[W]$. As $|X_{i'} \setminus Z| \leq |X_{i'}| - (\ell + 1)$, this transformation does not increase the width. □

## C  Safeness of the almost simplicial vertex rule

**Lemma 5.** *Let $G = (V, E)$ be a graph and let $v \in V$ be an almost simplicial vertex of degree at least three, with special neighbor $w$. Let $G'$ be obtained by deleting $v$ and by adding a vertex $v_{p,q}$ with neighbors $p$ and $q$ for any $p, q \in N_G(v)$ with $p \neq q$. Then $\mathbf{pw}(G) = \mathbf{pw}(G')$.*

*Proof.* Let $C := N(v) \setminus \{w\}$ and $C_v = C \cup \{v\}$. For ease of reading we denote by $v_{p,q}$ those vertices where $p, q \in C$ and by $v_{p,w}$ those adjacent to $w$ as well as to some $p \in C$.

($\geq$): We first show that $\mathbf{pw}(G) \geq \mathbf{pw}(G')$. Let $\mathcal{P} = (X_1, \ldots, X_r)$ be a path decomposition of $G$. We show how to get a path decomposition for $G'$ of at most the same width. Clearly $C_v = C \cup \{v\}$ is a clique in $G$, so there must be bags of $\mathcal{P}$ that completely contain $C_v$; we call those the $C_v$-bags (they must be connected).

If $w$ is contained in a $C_v$-bag, then we may delete $v$ from all other bags while maintaining a path decomposition for $G$. We copy this bag, such that we get a consecutive chain of $\binom{|N(v)|}{2}$ bags, one for each vertex $v_{p,q}$ and $v_{p,w}$. In each of those bags, we replace $v$ by a different $v_{p,q}$ or $v_{p,w}$ vertex. It is easy to see that this gives a path decomposition for $G'$ of at most the same pathwidth.

It remains to consider the more difficult case that $w$ is not contained in any $C_v$ bag. W.l.o.g., let $w$ occur only left of the $C_v$-bags and let $B$ denote the rightmost bag containing $w$. It follows that $v$ occurs in all bags from $B$ to the $C_v$ bags, in order to represent the edge $\{v,w\}$ and adhere to connectivity of occurrences. Let $B'$ be the bag directly on the left of the $C_v$ bags (possibly $B = B'$). It contains $v$, so there must be some $\hat{p} \in C$ which is not contained in $B'$, otherwise it would be a $C_v$-bag. (Note that this means that $w$ cannot be adjacent to $\hat{p}$ in $G$.) Further, let $B''$ denote the leftmost $C_v$-bag.

We make a small modification of $\mathcal{P}$ to prepare the replacements that are necessary to obtain a path decomposition for $G'$: We replace in $B'$ and all bags left of it the vertex $v$ by the vertex $w$, while we simply delete $v$ if $w$ is already present. At this point we have a path decomposition for $G$ except for not representing the edge $\{v,w\}$. The key observation is that $v$ and $w$ share the bag $B$ before these changes are made, and that all other neighbors of $v$ are in $C$ (and hence contained in the $C_v$-bag $B''$).

Now we will make replacements around the adjacent bags $B'$ and $B''$ in such a way that we obtain a path decomposition for $G'$. For ease of presentation let $B' = X' \cup \{w\}$ with $\hat{p} \notin X'$ and let $B'' = C \cup X \cup \{v\}$, where $X$ and $X'$ are vertex sets. We replace $B'$ and $B''$ by the following sequence of bags:

1. A copy of $B' = X' \cup \{w\}$.
2. Bags $\hat{C} \cup X \cup \{w\} \cup \{v_{p,w}\}$, one for each vertex $v_{p,w}$ with $p \in C$, with the rightmost being $\hat{C} \cup X \cup \{w\} \cup \{v_{\hat{p},w}\}$, where $\hat{C} := C \setminus \{\hat{p}\}$.
3. A bag $C \cup X \cup \{v_{\hat{p},w}\}$.
4. Bags $C \cup X \cup \{v_{p,q}\}$, one for each vertex $v_{p,q}$ with $p, q \in C$.

It is easy to see that all edges incident with vertices $v_{p,q}$ or $v_{p,w}$ are properly represented. It can also be verified that connectivity is not violated. The key points for this are that $\hat{p} \notin X'$, and that $B'$ and $B''$ were adjacent bags before the transformation. Thus after deleting all remaining occurrences of $v$, we obtain a path decomposition for $G'$ of at most the same width.

($\leq$): Now we show that $\mathbf{pw}(G) \leq \mathbf{pw}(G')$. Let $\mathcal{P}' = (X'_1, \ldots, X'_r)$ be a path decomposition of $G'$. We consider the bags of $\mathcal{P}'$ that completely contain $C$, the $C$-bags. If there is a $v_{p,w}$ vertex in a $C$-bag, then replacing all occurrences of that vertex by $v$, and deleting all other introduced vertices, we obtain a path decomposition for $G$: The reason is that $v_{p,w}$ occurs together with $C$, but also together with $w$. In this case we are done.

Otherwise, if the $C$-bags contain no $v_{p,w}$ vertex, then this implies that:

1. All vertices of $C$ must occur also outside of $C$-bags.
2. The vertex $w$ must occur at least once outside the $C$-bags.

Assume for contradiction that $w$ does not occur right of the $C$-bags. This would imply that all $v_{p,w}$ vertices can only be contained in bags left of the $C$-bags. Then however, all vertices $p \in C$ must occur left of the $C$-bags, contradicting the fact that only the $C$-bags contain $C$ completely.

Thus, by symmetry, $w$ occurs left and right of the $C$-bags. Hence, by connectivity of occurrences, $w$ is contained in all $C$-bags. It follows, that if any $C$-bags contains a $v_{p,q}$ vertex, we may replace that vertex by $v$, and again obtain a path decomposition for $G$ of the same width; observe that the bag in question contains $N(v) = C \cup \{w\}$. We will show that there is always a $C$-bag containing some $v_{p,q}$ vertex.

Assume for contradiction that no $C$-bag contains any $v_{p,q}$ vertex. We already know that each vertex of $C$ must occur outside of the $C$-bags, and we have used that not all those vertices can occur left of the $C$-bags (or not all right of them). Hence, there are distinct vertices $\hat{p}, \hat{q} \in C$ such that $\hat{p}$ does not occur right of the $C$-bags, and $\hat{q}$ does not occur left of them. Now, considering the vertex $v_{\hat{p},\hat{q}}$ this leads to a contradiction, as that vertex must occur both left and right of the $C$-bags, but by assumption it is not contained in any $C$-bag, contradicting connectivity of occurrences. □

*Remark 1.* Lemma 5 has a much simpler proof for the special case of $v$ being simplicial. It is easy to see that the replacement of $v$ does not generalize to the case of more special neighbors: Consider the 3-claw which has pathwidth one. Making an analogous replacement for the center of the claw, would create a cycle of length six which has pathwidth two. In general, when there is more than one special neighbor, then the modification creates a larger clique minor, than what the degree of $v$ would imply.

*Remark 2.* It can be easily seen that even the special case of simplicial vertices is not correct for treewidth. Application on one vertex of a clique of size four (and treewidth three) gives a graph of treewidth two.

## D  Hardness for Pathwidth and Treewidth Parameterized by a Modulator to a Single Clique

In this section, we give the proof that Treewidth and Pathwidth, parameterized by a modulator to a single clique do not have a polynomial kernel unless $\text{NP} \subseteq \text{coNP/poly}$. More precisely, we show this when we additionally restrict our input to co-bipartite graphs. As the treewidth equals the pathwidth for these graphs, the same proof can be used for the pathwidth as well as the treewidth version.

A *tree decomposition* of a graph $G = (V, E)$ is a pair $(\{X_i \mid i \in I\}, T = (I, F))$ with $\{X_i \mid i \in I\}$ a family of subsets of $V$, and $T$ a tree on edge set $F$, such that

– $\bigcup_{i \in I} X_i = V$.

- For all $\{v,w\} \in E$, there is an $i \in I$ with $v, w \in X_i$.
- For all $v \in V$, the set $I_v = \{i \in I \mid v \in X_i\}$ induces a subtree of $T$.

The sets $X_i$ are called the *bags* of the tree decomposition. The *width* of a tree decomposition $(\{X_i \mid i \in I\}, T = (I, F))$ is $\max_{i \in I} |X_i| - 1$, and the *treewidth* of $G$ is the minimum width of a tree decomposition of $G$.

If we have a weight function $w \colon V(G) \to \mathbb{N}$ then the *weighted width* of a tree decomposition $(\{X_i \mid i \in I\}, T = (I, F))$ of $G$ equals $\max_{i \in I} \sum_{v \in X_i} w(i)$, and the *weighted treewidth* of $G$ is the minimum weighted width of a tree decomposition of $G$. Observe that, contrary to the case of normal treewidth, we do not subtract one; hence the weighted treewidth of a tree in which $w(v) = 1$ for all vertices is two, rather than one.

An alternative characterization of treewidth is with help of elimination orderings. An *elimination ordering* of a graph is a permutation of its vertices. To eliminate a vertex $v$ corresponds to making its neighbors into a clique and then deleting $v$. Given an elimination ordering $\pi$ of a graph $G$, we obtain a sequence of graphs by eliminating the vertices of $G$ in the order of $\pi$. The following proposition is well known, see e.g., [3].

**Proposition 1.** *Graph $G$ has treewidth at most $k$ if and only if $G$ has an elimination ordering in which each vertex has degree at most $k$ when it is eliminated.*

The kernel lower bound is proven using the framework of cross-composition, which relies on the following notions.

**Definition 1 (Polynomial equivalence relation [6]).** *An equivalence relation $\mathcal{R}$ on $\Sigma^*$ is called a* polynomial equivalence relation *if the following two conditions hold:*

1. *There is an algorithm that given two strings $x, y \in \Sigma^*$ decides whether $x$ and $y$ belong to the same equivalence class in $(|x| + |y|)^{\mathcal{O}(1)}$ time.*
2. *For any finite set $S \subseteq \Sigma^*$ the equivalence relation $\mathcal{R}$ partitions the elements of $S$ into at most $(\max_{x \in S} |x|)^{\mathcal{O}(1)}$ classes.*

**Definition 2 (Cross-composition [6]).** *Let $L \subseteq \Sigma^*$ be a set and let $Q \subseteq \Sigma^* \times \mathbb{N}$ be a parameterized problem. We say that $L$* cross-composes *into $Q$ if there is a polynomial equivalence relation $\mathcal{R}$ and an algorithm which, given $t$ strings $x_1, x_2, \ldots, x_t$ belonging to the same equivalence class of $\mathcal{R}$, computes an instance $(x^*, k^*) \in \Sigma^* \times \mathbb{N}$ in time polynomial in $\sum_{i=1}^{t} |x_i|$ such that:*

1. $(x^*, k^*) \in Q \Leftrightarrow x_i \in L$ *for some $1 \leq i \leq t$,*
2. $k^*$ *is bounded by a polynomial in $\max_{i=1}^{t} |x_i| + \log t$.*

**Theorem 3 ([6]).** *If some set $L$ is NP-hard under Karp reductions and $L$ cross-composes into the parameterized problem $Q$ then there is no polynomial kernel for $Q$ unless $NP \subseteq coNP/poly$.*

We start with a number of results on (weighted) treewidth that are folklore or can easily be proved with standard arguments (see, e.g., [17]).

**Definition 3.** *Consider a graph G weighted by function w, and an elimination ordering $\pi$ on the vertices of G. The* cost *of $\pi$ is the maximum over all vertices $v \in V(G)$ of the weight of $N[v]$ at the time that $v$ is eliminated by $\pi$.*

**Proposition 2.** *Graph G has weighted treewidth at most k if and only if there is an elimination ordering of G of cost at most k.*

**Proposition 3 (Cf. [1]).** *Let G be a co-bipartite graph on bipartition $V(G) := A \dot\cup B$ weighted by function w. For every elimination ordering $\pi$ on G there is an elimination ordering $\pi'$ which does not cost more than $\pi$, such that $\pi$ first eliminates all vertices of A, and finishes by eliminating all vertices of B.*

**Proposition 4.** *Let G be a graph with weight function w containing two adjacent vertices $v, w$ such that $N_G[v] \subseteq N_G[w]$. Let $\pi$ be an elimination ordering of G which eliminates $w$ before $v$, and let the ordering $\pi'$ be obtained by updating $\pi$ such that it eliminates $v$ just before $w$. Then the cost of $\pi'$ is not higher than the cost of $\pi$.*

**Proposition 5.** *Let G be a graph with positive integral vertex weights. Let $G'$ be the graph obtained from G by iterating the following procedure. As long as there is a vertex $v$ with weight more than one, subtract one from the weight of $v$ and add a new vertex $v'$ of weight 1 and with neighborhood $N[v]$. Then the treewidth of $G'$ equals the weighted treewidth of $G'$ minus one.*

**Theorem 4.** TREEWIDTH PARAMETERIZED BY A MODULATOR TO A SINGLE CLIQUE *does not admit a polynomial kernelization unless $NP \subseteq coNP/poly$.*

*Proof.* We show that the NP-complete CUTWIDTH3 problem cross-composes into TWMSC. We start by defining a polynomial equivalence relationship $\mathcal{R}$. Fix an encoding of instances of CUTWIDTH3, and choose $\mathcal{R}$ such that all strings which do not encode a valid instance are equivalent. For the strings which *do* encode a valid instance, define two instances $(G_1, k_1)$ and $(G_2, k_2)$ to be equivalent if all of the following hold: $k_1 = k_2$, $|V(G_1)| = |V(G_2)|, |E(G_1)| = |E(G_2)|$, and for each integer $i \in \{1, 2, 3\}$ the number of degree-$i$ vertices in $G_1$ and $G_2$ is the same. Since a set of valid instances on at most $n$ vertices each is partitioned into at most $n \times n \times n^3$ equivalence classes, this constitutes a polynomial equivalence relationship.

We now show how to cross-compose a set of instances of CUTWIDTH3 which belong to the same equivalence class of $\mathcal{R}$. If all instances are malformed, then this can be recognized in polynomial time and we simply output a constant-size **no**-instance. So in the remainder we may assume that all input instances $(G_1, k_1)$, $\ldots, (G_t, k_t)$ are well-formed and belong to the same equivalence class; in particular $k_1 = \ldots = k_t = k$ and $|V(G_1)| = \ldots = |V(G_t)| = n$. Order the vertices within each graph by increasing degree, breaking ties arbitrarily. The choice of $\mathcal{R}$, together with the fact that each $G_i$ has maximum degree 3, guarantees that each graph has the same number of vertices of each degree.

Since CUTWIDTH on a graph on $n$ vertices can be solved in $\mathcal{O}^*(2^n)$ time [5, Theorem 10], we may assume that $n \geq \log t$. For if $n < \log t$ then applying the algorithm by Bodlaender et al. [5] consecutively on each instance can be done in time which is polynomial in the total input size (which is at least $t$); we could then output a constant-size instance with the appropriate answer as the output of the cross-composition. For similar reasons we may assume $n \geq 2$. Finally, we may assume that the number of input instances $t$ is a power of 2 since we can duplicate some instances without changing the value of the OR, increasing the input size by at most a factor two.

To construct the instance of TWMSC that encodes the OR of the input instances, we use a two-stage process for the ease of presentation. We first show that the OR of the input instances can be encoded into an instance of WEIGHTED TREEWIDTH PARAMETERIZED BY A MODULATOR TO A SINGLE CLIQUE on a co-bipartite graph with partite sets $A$ and $B$, such that the total weight of the set $B$ is polynomial in $n + \log t$. The set $B$ will be the modulator, which is valid since removing the partite set $B$ from a co-bipartite graph leaves a clique. We then use Proposition 5 to obtain an equivalent instance of TWMSC, and since the total weight of $B$ is sufficiently small this produces an instance of TWMSC that encodes the OR of the input instances, and has a modulator to a single clique whose size is polynomial in $n + \log t$.

We now construct a graph $G^*$ and weight function $w$ such that computing the weighted treewidth of $G^*$ corresponds to computing the OR of the instances of CUTWIDTH3. The construction is based on the NP-completeness proof for TREEWIDTH by Arnborg et al. [1]. The graph $G^*$ will be co-bipartite with partite sets $A$ and $B$, so $V(G^*) := A \dot\cup B$ and $A$ and $B$ are cliques in $G^*$. The graph $G^*$ is defined as follows:

- For each input graph $G_i$ with $i \in [t]$, for each vertex $j \in V(G_i)$, we add a vertex $v_{i,j}$ of weight $n^3$ to $A$ which corresponds to vertex $j$. For a given value of $j \in [n]$ we say that all vertices $v_{i,j}$ (for all relevant values of $i$) are *A-representatives* of node $j$. We also add a *dummy* vertex $d_i$ for each instance $i \in [t]$ to $A$ of weight $n^6$. We turn $A$ into a clique.
- The vertex set $B$ consists of three parts: the *instance selector vertices* $B_I$, the *node representatives* $B_N$ and the *edge representatives* $B_E$.
  - The instance selector vertices will be used to encode the binary representation of instance numbers. Since we assumed $t$ to be a power of two, we need $\log t$ bits to encode an instance number and therefore $2 \log t$ vertices are used to represent all possible bit values for $\log t$ positions. So $B_I := \{a_q, b_q \mid q \in [\log t]\}$. Each vertex in $B_I$ has weight $n^5$.
    We connect the vertices of $B_I$ to the vertices of $A$ as follows. We make a vertex $v_{i,j}$ in $A$ (which corresponds to instance $i$) adjacent to the instance selector vertices of the bit values of the binary representation of $i$. So for $q \in [\log t]$, if the $q$-th bit of number $i$ is 1 then we make $v_{i,j}$ adjacent to $a_q$, and if the bit is 0 then we make the vertex adjacent to $b_q$. The adjacency from dummy vertices $d_i$ to the vertices of $B_I$ is defined exactly the same through the binary representation of $i$.
  - The node representatives $B_N$ contain a vertex for each node number in $[n]$. Recall that all input graphs have the same number of vertices of each degree, and that we sorted the vertices by degree. When we write $\deg(j)$ for $j \in [n]$ we will therefore take this to mean the value $d$ such that in each input graph, each vertex $j$ has degree $d$. For each $j \in [n]$ we add a vertex $x_j$ to the set $B_N$ and give it weight $n^3 - \deg(j)$. The vertex $x_j$ is said to be the (unique) *B-representative* of node $j$.

The adjacency between $B_N$ and $A$ is simple: for each $j \in [n]$ we make all $A$-representatives of $j$ adjacent to the single $B$-representative of $j$, and we make all the nodes in $B_N$ adjacent to all the dummy vertices $d_i$ for $i \in [t]$.

- The edge representatives $B_E$ contain one vertex for each possible edge in an undirected $n$-vertex graph. So for $\{v,w\} \in \binom{[n]}{2}$ we have a vertex $e_{v,w}$ of weight two. Vertex $e_{v,w}$ is adjacent to an $A$-representative $v_{i,j}$ if instance $G_i$ contains the edge $\{v,w\}$ and $j = v$ or $j = w$, i.e., the edge representative $e_{v,w}$ is adjacent to instance $i$'s $A$-representatives of the endpoints of the edge, provided that instance $i$ actually contains the edge. Additionally, all vertices of $B_E$ are adjacent to all dummy vertices $d_i$ for $i \in [t]$.

The construction is completed by turning $B := B_I \cup B_N \cup B_E$ into a clique. We set $k' := t \cdot (n^4 + n^6) + n^3 + n^5 \log t + k$.

To complete the first stage, we need to prove that $(G^*, w)$ has weighted treewidth at most $k'$ if and only if at least one of the input graphs $G_i$ has cutwidth at most $k$. Before proving this claim, we establish some properties of the constructed instance $(G^*, w, k')$.

*Claim 1.* Let $S := \{v_{i,j} \mid j \in [n]\}$ for a given instance number $i \in [t]$ be the subset of the vertices in $A$ corresponding to instance $i$. Let $\pi$ be a permutation of $S$. Consider the process of eliminating the vertices in $S$ from graph $G^*$ in the order given by $\pi$, and let E-WEIGHT$(S_{\pi(j)})$ be the total weight of $N[S_{\pi(j)}]$ when eliminating the vertex $\pi(j)$ for $j \in [n]$. Then E-WEIGHT$(S_{\pi(j)}) = t \cdot (n^4 + n^6) + n^3 + n^5 \log t + \ell$, where $\ell := |\{\{u,v\} \in E(G_i) \mid \pi(u) \leq j < \pi(v)\}|$.

*Proof.* The intuition behind the proof is that the elimination process has two effects on the weight of neighbors of some vertex $v \in A$: on the one hand, eliminated vertices in $A$ are essentially replaced by the representatives in $B_N$ in the neighborhood of $v$, which have slightly smaller weight than the originals; the difference is exactly equal to the degree of the corresponding vertex. On the other hand, the representative of any edge in $B_E$ will be added to those neighborhoods, once one of the endpoints is eliminated; recall that those edges contribute a weight of two. Thus, when reaching the first endpoint of an edge, the weight increases by one (by the degree contribution); when reaching the second endpoint this increase is canceled. Together this leads to the contribution of $\ell$ in E-WEIGHT$(S_{\pi(j)})$. This idea was used by Arnborg et al. [1] in their NP-completeness proof for TREEWIDTH.

Armed with this intuition, let us proceed with the proof. By definition of $G^*$, all vertices in $S$ have the same set of neighbors in $B_I$ so elimination of vertices from $S$ does not affect the adjacency of other vertices in $S$ to $B_I$. Consider a vertex $v_{i,j}$ in $S$. From the construction of $G^*$ it follows that initially, the only vertex of $S$ which is adjacent to the $B$-representative of $j$, is the vertex $v_{i,j}$. Since we only eliminate vertices from $S$, it follows that a vertex in $S$ is only adjacent to the $B$-representative of a node number $j$ if that vertex is itself the unique $A$-representative of $j$ in $S$, or if the $A$-representative of $j$ in $S$ was eliminated earlier. Let us use these observations to prove the claim.

For an arbitrary value of $j \in [n]$ we consider the closed neighborhood of the vertex $S_{\pi(j)}$ just before it is eliminated. We will study the neighborhood of $S_{\pi(j)}$ in the sets $A, B_I, B_N$ and $B_E$ consecutively. For convenience, define E-WEIGHT$^X(S_{\pi(j)})$ for $X \subseteq V(G^*)$ as the total weight of $N[S_{\pi(j)}] \cap X$ when $S_{\pi(j)}$ is eliminated.

**Neighbors in $A$.** Since $A$ is a clique and $S \subseteq A$, vertex $S_{\pi(j)}$ is initially adjacent to all vertices of $A$. Since the only vertices which are eliminated are those in $S$ corresponding to instance $i$, vertex $S_{\pi(j)}$ will be adjacent to all vertices for other instances, i.e., to $v_{i',j}$ for $i' \neq i$ and $j \in [n]$, for a total weight of E-WEIGHT$^{A\setminus S}(S_{\pi(j)}) = (t-1)n \cdot n^3$. Vertex $S_{\pi(j)}$ is also adjacent to all $t$ dummy vertices for a weight of $n^6$ each. The remaining vertices of $A$ are those in $S$, and $S_{\pi(j)}$ is adjacent to those which are not already eliminated. Hence there are $n - j + 1$ vertices in $S$ which are in the closed neighborhood of $S_{\pi(j)}$ just before it is eliminated. These have weight $n^3(n - j + 1)$ so E-WEIGHT$^A(S_{\pi(j)}) = (t-1)n^4 + t \cdot n^6 + n^3(n - j + 1)$.

**Neighbors in $B_I$.** Since the neighborhood of $S_{\pi(j)}$ in $B_I$ is not changed by the eliminations, vertex $S_{\pi(j)}$ has exactly $\log t$ neighbors in $B_I$ with weight $n^5$ each so E-WEIGHT$^{B_I}(S_{\pi(j)}) = n^5 \log t$.

**Neighbors in $B_N$.** By construction of $G^*$, vertex $S_{\pi(j)}$ is adjacent to the unique node in $B_N$ which is the $B$-representative for the vertex for which $S_{\pi(j)}$ is an $A$-representative. Initially, $S_{\pi(j)}$ is not adjacent to other vertices of $B_N$. For each vertex $1 \leq j' < j$ which was eliminated before $j$, vertex $S_{\pi(j)}$ has become adjacent to the vertex in $B_N$ which is $B$-representative for the vertex to which $S_{\pi(j')}$ is the $A$-representative. So E-WEIGHT$^{B_N}(S_{\pi(j)}) = \sum_{j'=1}^{j}(n^3 - \deg(j'))$.

**Neighbors in $B_E$.** Initially, vertex $S_{\pi(j)}$ is adjacent to the edge-representative vertices in $B_E$ for which $S_{\pi(j)}$ represents an endpoint, so to $\deg(j)$ vertices with weight two each. For each vertex $S_{\pi(j')}$ with $1 \leq j' < j$ which is eliminated before $\pi(j)$, $S_{\pi(j)}$ becomes adjacent to the edge-representatives in $B_E$ for edges which are incident on $S_{\pi(j')}$ in graph $G_i$. This shows that E-WEIGHT$^{B_E}(S_{\pi(j)}) := 2|\bigcup_{1 \leq j' \leq j}\{e_{u,v} \mid \{u,v\} \in E(G_i) \land (j' = u \lor j' = v)\}|$.

We can now sum up the weights of the members of the closed neighborhood $N[S_{\pi(j)}]$ in each of the respective subsets to establish that E-WEIGHT$(S_{\pi(j)})$

equals:

$$\overset{A}{\text{E-WEIGHT}}(S_{\pi(j)}) + \overset{B_I \cup B_N \cup B_E}{\text{E-WEIGHT}}(S_{\pi(j)})$$

$$=[(t-1)n^4 + t \cdot n^6 + n^3(n-j+1)] + [n^5 \log t] + \left[\sum_{j'=1}^{j}(n^3 - \deg(j'))\right] +$$

$$2\left|\bigcup_{1 \leq j' \leq j}\{e_{u,v} \mid \{u,v\} \in E(G_i) \wedge (j' = u \vee j' = v)\}\right|$$

$$=t \cdot (n^4 + n^6) + n^3 + n^5 \log t - \sum_{j'=1}^{j}(\deg(j'))+$$

$$2\left|\bigcup_{1 \leq j' \leq j}\{e_{u,v} \mid \{u,v\} \in E(G_i) \wedge (j' = u \vee j' = v)\}\right|$$

To simplify this further, we define $E_1$ as the set of edges of $G_i$ which have one endpoint among the vertices in the range $[1 \ldots j]$, and $E_2$ as the edges of $G_i$ with both endpoints among $[1 \ldots j]$. Observe that these definitions imply that $\sum_{j'=1}^{j} \deg(j') = |E_1| + 2|E_2|$ and $|\bigcup_{1 \leq j' \leq j}\{e_{u,v} \mid \{u,v\} \in E(G_i) \wedge (j' = u \vee j' = v)\}| = |E_1| + |E_2|$. We continue the derivation:

$$=t \cdot (n^4 + n^6) + n^3 + n^5 \log t - (|E_1| + 2|E_2|) + 2(|E_1| + |E_2|)$$
$$=t \cdot (n^4 + n^6) + n^3 + n^5 \log t + |E_1|.$$

Now observe that by definition, $|E_1|$ is the number of edges which have one endpoint at or to the left of $j$, and the other endpoint to the right of $j$, and hence this is exactly the value of $\ell$ as defined in the statement of the claim; this concludes the proof of Claim 1. □

The preceding claim relates the cost of the first $n$ eliminations of an ordering of $G^*$ to the cutwidth of an instance $i$, provided that the ordering starts by eliminating the $A$-representatives of instance $i$. The next claim shows that these first $n$ eliminations essentially dominate the cost of elimination orderings with this structure.

*Claim 2.* Let $S$, $i$, $\pi$ and E-WEIGHT be as defined in Claim 1. Consider an elimination ordering for $G^*$ which starts by eliminating $S$ in the order given by $\pi$, then eliminates the dummy $d_i$ corresponding to instance $i$, and eliminates the remaining vertices in arbitrary order. The cost of $\pi^*$ is $\max_{j \in [n]}$ E-WEIGHT$(S_{\pi(j)})$.

*Proof.* By Claim 1, the maximum weight of a closed neighborhood when eliminating the vertices from $S$ is exactly $\max_{j \in [n]}$ E-WEIGHT$(S_{\pi(j)}) \geq t \cdot (n^4 + n^6) + n^3 + n^5 \log t$. We show that after elimination of $S$, eliminating the dummy $d_i$ and all remaining vertices does not incur a cost higher than this.

Consider the weight of the closed neighborhood of the dummy vertex $d_i$ after the $n$ vertices from $S$ have been eliminated. At that stage, $d_i$ is adjacent to all vertices which are left in $A$, to all vertices of $B_N$, some vertices of $B_E$, and to the $\log t$ vertices in $B_I$ which correspond to the binary representation of the number $i$. Since the total weight of $B_N$ is not more than $n \cdot n^3$, the weight of $N[d_i]$ when $d_i$ is eliminated is bounded by $t(n^4+n^6) - n^4 + n \cdot n^3 + n^5 \log t + 2\binom{n}{2}$, which is at most $t(n^4+n^6) + n^3 + n^5 \log t$ and so does not exceed the cost incurred for the first $n$ vertices.

After $d_i$ and $S$ have been eliminated from the graph, the total weight of the remaining vertices is at most $(t-1)(n^4+n^6) + 2n^5 \log t + n^4 + 2\binom{n}{2}$, which is bounded by $t(n^4+n^6) + n^3 + n^5 \log t$ as we assumed $n \geq \log t$ at the beginning of the proof. Hence the cost of this elimination ordering $\pi^*$ is dominated by the cost of eliminating the first $n$ vertices and is therefore $\max_{j \in [n]}$ E-WEIGHT$(S_{\pi(j)})$. □

Having shown how the cost of specific types of elimination orderings of $G^*$ corresponds to the cutwidth of one particular input instance, we proceed to show that there is always an optimal ordering of this type.

*Claim 3.* If there is an elimination ordering of $(G^*, w)$ of cost at most $k'$, then there is such an ordering which starts by eliminating all vertices in the set $\{v_{i,j} \mid j \in [n]\}$ for some $i \in [t]$, i.e., there is an ordering which first eliminates all A-representatives corresponding to one particular input instance $G_i$.

*Proof.* The proof contains of two parts. We first give a canonical elimination ordering of bounded cost, and then use a replacement argument which compares the cost of this canonical ordering to an ordering which does not match the form in the statement of the lemma.

Define a canonical elimination $\pi^*$ as follows. It starts with the sequence $v_{1,1}, v_{1,2}, \ldots, v_{1,n}$, then eliminates dummy $d_1$, and finally eliminates the rest of the vertices in arbitrary order. By Claim 2 the cost incurred by ordering $\pi^*$ is $\max_{j \in [n]}$ E-WEIGHT$(S_{\pi(j)})$. Observe that the cutwidth of graph $G_1$ under any ordering does not exceed the number of its edges, which is at most $3n/2$ since $G_1$ has maximum degree at most three. Hence we find that the term $\ell$ in the expression for E-WEIGHT given by Claim 1 is bounded by $3n/2$. Using Claim 1 we therefore find that the cost of this canonical elimination ordering $\pi^*$ is bounded by $t \cdot (n^4+n^6) + n^3 + n^5 \log t + 3n/2$.

To complete the proof, the second part will show that any elimination ordering whose form does not match that in the statement of the claim, has cost as least as much as the canonical ordering. So consider any elimination ordering $\pi$ of $(G^*, w)$ of cost at most $k'$. By Proposition 3 there is an optimal-cost elimination ordering which first eliminates all of $A$, so assume without loss of generality that $\pi$ first eliminates all vertices of $A$. Since the construction of $G^*$ guarantees that for all $i \in [t]$, all vertices $v_{i,j}$ of $A$ satisfy $N_{G^*}[v_{i,j}] \subseteq N_{G^*}[d_i]$, Proposition 4 shows that we may assume without loss of generality that for all $i \in [t]$, the vertices $v_{i,j}$ for $j \in [n]$ are eliminated by $\pi$ earlier than $d_i$; hence the first $n$ vertices eliminated by $\pi$ are not dummy vertices. If the first $n$ vertices correspond to the same instance then we are done; hence in the remainder we may assume this is not the case. Consider the first index $1 < j \leq n$ such that all vertices $\pi(j')$ for $1 \leq j' < j$ correspond to the same instance $i$ (i.e., they are of the form $v_{i,j''}$ for $j'' \in [n]$) and $\pi(j)$ corresponds to instance $i'$ with $i \neq i'$. Let us consider the neighborhood of the vertex $\pi(j)$ when it is eliminated.

By construction of $G^*$, vertex $\pi(j)$ corresponding to instance $i'$ is adjacent to the vertices in $B_I$ which correspond to the binary representation of $i'$. Since vertex $\pi(1)$ was eliminated before $\pi(j)$, and since vertices $\pi(1)$ and $\pi(j)$ are adjacent in $G^*$ because they are both members of the clique $A$, after elimination of $\pi(1)$ the vertex $\pi(j)$ has become adjacent to all neighbors of $\pi(1)$. Since $\pi(1)$ is adjacent to the vertices of $B_I$ corresponding to the binary representation of $i$, and since the binary representations of $i$ and $i'$ must differ in at least one position, the number of neighbors of $\pi(j)$ in $B_I$ at the time it is eliminated is at least $1 + \log t$, and they have weight $n^5$ each. Since $\pi(j)$ is also adjacent to all vertices of $A$ except the $j-1$ vertices of weight $n^3$ which were eliminated earlier, this shows that the weight of the closed neighborhood of $\pi(j)$ at the time it is eliminated is at least $t(n^4 + n^6) - j \cdot n^3 + (1 + \log t)n^5$. Using that $j \leq n$ and $n \geq 2$ (which we assumed in the beginning the proof of the theorem), it now follows that the weight of $\pi(j)$ at the time it is eliminated is at least as much as the cost of the canonical elimination ordering. Hence the canonical elimination ordering which we defined earlier has cost no more than $\pi$, and since the canonical ordering starts by eliminating $v_{1,1}, v_{1,2}, \ldots, v_{1,n}$ this concludes the proof of Claim 3.    □

We are now finally ready to prove that $(G^*, w)$ has weighted treewidth at most $k'$ if and only if at least one of the input graphs $G_i$ has cutwidth at most $k$. First assume that $(G^*, w)$ has weighted treewidth at most $k'$. By Proposition 2 this implies that there is an elimination ordering $\pi$ of $G^*$ with cost at most $k'$. By Claim 3 we may assume that there is an instance number $i^* \in [t]$ such that $\pi$ starts by eliminating all vertices in the set $S := \{v_{i^*,j} \mid j \in [n]\}$. As the cost of $\pi$ is at most $k'$, the weight of the closed neighborhood of a vertex in $S$ at the time it is eliminated does not exceed $k'$. By Claim 1 this proves that $\max_{j \in [n]} \text{E-WEIGHT}(S_{\pi(j)}) \leq k'$. Plugging in the value for $k'$ and the expression for E-WEIGHT obtained in the mentioned claim, and cancelling terms on both sides, we find that $\max_{j \in [n]} |\{\{u, v\} \in E(G_{i^*}) \mid \pi(u) \leq j < \pi(v)\}| \leq k$ which proves that $G_{i^*}$ has cutwidth at most $k$, when using the ordering on $S$ induced by $\pi$.

For the reverse direction, assume that $G_{i^*}$ has cutwidth at most $k$, and let $\pi^*$ be an ordering which achieves this cutwidth. Build an elimination ordering for $G^*$ by first eliminating the vertices of $S := \{v_{i^*,j} \mid j \in [n]\}$ in the order induced by $\pi^*$, then eliminating the dummy $d_{i^*}$, and then eliminating the remaining vertices in arbitrary order. By Claim 2 the cost of this ordering is dominated by the cost of eliminating the vertices of $S$, which is $\max_{j \in [n]} \text{E-WEIGHT}(S_{\pi(j)})$. If ordering $\pi^*$ achieves cutwidth at most $k$ on $G_{i^*}$, then evaluating the expression for E-WEIGHT given by Claim 1 proves that the cost of $\pi$ is at most $k'$. Using Proposition 2 this proves that $(G^*, w)$ has weighted treewidth at most $k'$.

To complete the cross-composition of CUTWIDTH3 into TWMSC, we can transform the weighted graph $(G^*, w)$ to the unweighted graph $\hat{G}$ using the transformation of Proposition 5. Since this transformation duplicates the closed neighborhoods of vertices, it results in a co-bipartite graph since the cliques $A$ and $B$ of $G^*$ are just transformed into larger cliques in $\hat{G}$. Let $\hat{B}$ be the clique in $\hat{G}$ which results from the transformation of clique $B$ in $G^*$. The size of $\hat{B}$ is bounded by the maximum weight of a vertex in $B$ (under $w$) times the size of $B$. Since both are polynomial in $n + \log t$, this shows that the size of $\hat{B}$ is bounded polynomially in $n + \log t$. Now consider the instance of TWMSC which asks if $\hat{G}$ with the modulator $\hat{B}$ to a single clique (because $\hat{G}$ is co-bipartite and $\hat{B}$ is one of the partite sets) has treewidth at most $k' - 1$; by the equivalence between the weighted treewidth of the original graph, and the normal treewidth of the result of the transformation, our constructed instance is equivalent to the OR of the input instances of CUTWIDTH3. The size of the modulator, which is the parameter of the TWMSC instance, is polynomial in $n + \log t$. This concludes the cross-composition; Theorem 4 follows by applying Theorem 3.   □

Since the pathwidth of a co-bipartite graph equals its treewidth [14] and the graph formed by the cross-composition is co-bipartite, we obtain the following corollary.

**Corollary 1.** PATHWIDTH PARAMETERIZED BY A MODULATOR TO A SINGLE CLIQUE *does not admit a polynomial kernel unless* $NP \subseteq coNP/poly$.